\documentstyle[prl,twocolumn,aps,epsf,eqsecnum]{revtex}
\begin{document}

\hoffset-1cm


\title{Two-particle interferometry for non-central 
heavy-ion collisions}
\author{Urs Achim Wiedemann}

\address{
   Institut f\"ur Theoretische Physik, Universit\"at Regensburg,
   D-93040 Regensburg, Germany
}

\date{\today}
\maketitle

\begin{abstract}
In non-central heavy ion collisions, identical two particle
Hanbury-Brown/Twiss (HBT) correlations $C({\bf K},{\bf q})$
depend on the azimuthal direction of the pair
momentum ${\bf K}$. We investigate the consequences for a 
harmonic analysis of the corresponding HBT radius parameters
$R_{ij}^2$. Our discussion includes both, a model-independent 
analysis of these parameters in the Gaussian approximation, 
and the study of a class of hydrodynamical models which mimic 
essential geometrical and dynamical properties of peripheral 
heavy ion collisions. Also, we discuss the additional 
geometrical and dynamical information contained in the harmonic 
coefficients of $R_{ij}^2$. The leading contribution of their
first and second harmonics are found to satisfy simple constraints.
This allows for a minimal, azimuthally sensitive parametrization
of all first and second harmonic coefficients in terms of only
two additional fit parameters. We determine to what extent these 
parameters can be extracted from experimental data despite finite 
multiplicity fluctuations and the resulting uncertainty in the 
reconstruction of the reaction plane.
\end{abstract} 


\section{Introduction}
\label{sec1}
The goal of the current and future experimental heavy ion 
programs at CERN and BNL is to test the equilibration properties 
of hadronic matter at energy densities where quarks and gluons 
are the relevant physical degrees of freedom. Anisotropic 
(directed) transverse flow is an important observable
for this program, since both hydrodynamic and thermodynamic 
behaviour is based on equilibration processes between local
degrees of freedom. Hydrodynamical flow effects
result from pressure gradients which due to compression of the 
hadronic matter build up during the collision process.
Their strength depends on the equation of state of the hot
matter and provides insight into the collision dynamics.
Moreover, hadronic observables mainly test the final collision
stage and a back extrapolation is needed to extract from 
them information about the hot and dense earlier stages.
For this to work, collective and random ('thermal') motion 
in the collision region have to be distinguished properly. 
Hence, concepts of different types of collective flow play a 
central role in understanding the dynamics of heavy ion collisions. 

Directed anisotropic flow was observed in both AGS~\cite{B94,B97} and 
SPS~\cite{WIE95,SEY95} experiments, as well as at lower BEVALAC/SIS 
energies~\cite{GUT90}. Directivity\cite{WIE95,SEY95}, 2- and 3-dimensional 
sphericity~\cite{DG83,O92,O93}, or the so-called deformation parameter 
$R_p$~\cite{GUT90} are typical variables used in
its characterization. With minor differences, all of them are sensitive 
to azimuthal anisotropies in the triple-differential particle 
distributions. The most complete experimentally feasible parametrization
is obtained in a Fourier expansion in the azimuthal
angle for different values of rapidity and transverse 
momentum,~\cite{VZ96,B94,B97,V97} 
  \begin{eqnarray}
    E{dN\over d^3p} &=&
    {d^3N\over p_t\, dp_t\, dy\, d\phi} = \int d^4x S(x,p)
    \nonumber \\
    &=& \textstyle{1\over 2\pi} {d^2N\over p_t\, dp_t\, dy}
        \lbrack 1 + 2\sum_{n=1}^\infty v_n\cos n(\phi -\psi_R^{(n)}) 
                \rbrack\, .
    \label{1.1}
  \end{eqnarray}      
Here, the azimuthal angles $\psi_R^{(n)}$ allow for the determination
of the reaction plane and the harmonic coefficients $v_n$ characterize
the size of the total vector sum of transverse momenta ($n = 1$), the
approximate elliptic shape of the azimuthal distribution
($n = 2$) and higher order triangle-type ($n=3$), rectangle-type ($n=4$)
etc. deformations. 

In Eq. (\ref{1.1}), we have expressed the one-particle distribution in
terms of the emission function $S(x,p)$ which specifies the collision 
region at freeze-out. $S(x,p)$ is a Wigner distribution
and denotes the phase space probability that a particle of four momentum 
$p$ is emitted from a space-time point $x$ in the collision region. 
Features of collective dynamics as e.g. directed flow are encoded
in the source function $S(x,p)$ as $x$-$p$ position-momentum correlations.
Observables extracted from the one-particle
distributions (\ref{1.1}) are not sensitive to the space-time characterisitics
(and a fortiori to $x$-$p$ correlations) of the source. 
The question arises to what extent observables which 
are sensitive to $x$-$p$ correlations can support and refine the 
picture obtained via the analysis of (\ref{1.1}). This motivates
an azimuthally sensitive Hanbury-Brown Twiss analysis of 2-particle 
correlation functions, which is the main focus of the present work.

Identical two-particle correlations $C({\bf K},{\bf q})$, here written
in terms of the average $K = \textstyle{1\over 2}(p_1 + p_2)$ and
relative $q = p_1 - p_2$ pair momentum, are the only known observables
sensitive to space-time characteristics of the source. Their space-time 
interpretation is based on the result~\cite{S73,GKW79,P84,CH94} 
  \begin{eqnarray}
     C({\bf K},{\bf q})
      &=& 1 + \vert\langle e^{i{\bf q}{\cdot}({\bf x}-\vec{\beta}t)}
            \rangle\vert\, ,
            \nonumber \\
   \langle f(x) \rangle &=& {\int d^4x f(x) S(x,K)\over \int d^4x S(x,K)}\, ,
   \label{1.2}
  \end{eqnarray}
where we have used the on-shell condition $q\cdot K = 0$ to 
substitute the temporal component $q^0$ in the 4-dimensional Fourier
transform, $\vec{\beta} = {\bf K}/K_0$. According to (\ref{1.2}), 
determining $K$-dependent geometrical and dynamical source infomation 
reduces to a Fourier inversion problem (which due to the on-shell 
constraint however does not have a unique solution). 
In the standard analysis of (\ref{1.2}), one assumes an azimuthally
symmetric collision and characterizes $C({\bf K},{\bf q})$ with four
Gaussian Hanbury-Brown/Twiss HBT radius parameters which depend on 
$\vert {\bf K}_\perp\vert$ and the longitudinal pair rapidity $Y$ 
only~\cite{CSH95a}. 
In contrast, the anisotropic case requires six HBT radius 
parameters $R_{ij}$ which in addition depend on the azimuthal 
angle $\Phi$ of ${\bf K}_\perp$~\cite{H96}. 
Previous discussions of the azimuthal dependence of 
$C({\bf K},{\bf q})$ were based on event generator 
studies~\cite{VC96a,F96} for finite impact parameter collisions,
or exploited the Lorentz invariance of the correlator to derive 
azimuthally dependent HBT radius parameters~\cite{VC96b}.
The present work is complementary to these and starts in 
analogy to (\ref{1.1}) from expanding the angular dependence of
$C({\bf K},{\bf q})$ in a harmonic series,
  \begin{eqnarray}
    C({\bf K},{\bf q}) &=& 1 + 
         \exp{\left({ \sum_{ij} R_{ij}^2 q_iq_j}\right)}\, ,
         \nonumber \\
    R_{ij}^2(K_\perp,\Phi,Y) &=& 
             R_{ij,0}^2(K_\perp,Y)
               \nonumber \\
             &+& 
             2 \sum_{n=1}^\infty
               {R^{c}_{ij,n}}^2(K_\perp,Y) \cos n\Phi
               \nonumber \\
             &+& 
             2 \sum_{n=1}^\infty
               {R^{s}_{ij,n}}^2(K_\perp,Y) \sin n\Phi\, .
     \label{1.3}
  \end{eqnarray}
Here, the components $i$,$j$ are given in the ``out-side-long'' 
{\it osl}-system where the relative pair momentum has a transverse 
{\it out} component $q_o$ parallel to the  pair momentum 
${\bf K}_\perp$, a longitudinal {\it long} component $q_l$ along 
the beam and a remaining {\it side} component. 
To discuss the azimuthal $\Phi$-dependence of the two-particle 
correlator (\ref{1.3}), we frequently use an 
impact-parameter fixed system. In this system, the direction of
the impact parameter $\vec{b}$ specifies the $x$-axis,
the $z$-axis is along the beam and the $y$-axis is perpendicular
to the reaction plane spanned by $x$ and $z$. Accordingly, the total
angular momentum $\vec{L}$ of the system, with
  \begin{equation}
    L_i = \epsilon_{ijk} \langle\!\langle x_j\, p_k\, \rangle\!\rangle
        = \epsilon_{ijk} \int {d^3p\over E} \int d^4x\, x_j\, p_k\,
          S(x,p)\, ,
   \label{1.4}
  \end{equation}
points along the $y$ direction.

One central theme of the following is to determine those harmonic
coefficients ${R_{ij,n}^{c}}^2$, ${R_{ij,n}^{s}}^2$, whose 
contributions are not negligible. We shall find that there are 
very few independent ones. This makes a comparisons with experimental 
data feasible. Also, we 
aim at understanding which geometrical and dynamical information
about the particle emitting source is contained in the harmonic 
coefficients. In section~\ref{sec2}, we attack both problems
by deriving model-independent expressions for the HBT radius parameters.
These allow for the
calculation of HBT radii from $\Phi$-dependent space-time variances
$\langle x_\mu x_\nu\rangle$ of arbitrary model emission functions
$S(x,K)$. Investigating the $\Phi$-dependence of 
$\langle x_\mu x_\nu\rangle$ leads then to relations between the harmonic
coefficients in (\ref{1.3}). 
In the more detailed model-independent discussion in section~\ref{sec3}
and the subsequent model study in section~\ref{sec4}, we restrict 
our investigation to mid-rapidity and to symmetric collision systems.
Due to the reflection symmetry with respect to the $y$-$z$-plane,
all odd harmonic coefficients vanish in this case, and this 
considerably simplifies the discussion. In section~\ref{sec5},
we extend this analysis to the fragmentation regions. Again, we derive
simple relations between the non-vanishing first harmonic coefficients,
and we illustrate our findings quantitatively in a subsequent model
study. The discussion in sections~\ref{sec2}-\ref{sec5} implicitly
assumes that the orientation $\psi_R$ of the reaction plane is known.
It hence neglects finite multiplicity fluctuations which introduce
in practice a significant uncertainty in determining $\psi_R$. In 
section~\ref{sec6} we investigate to what extent information about
the anisotropy of the correlator $C({\bf K},{\bf q})$ can be obtained
despite these statistical constraints. The main
results are then summarized in the Conclusion.

\section{Azimuthal Dependence of Cartesian HBT Radius Parameters}
\label{sec2}

Here, we derive model-independent expressions for the
Cartesian HBT radius parameters (\ref{1.3}) in terms of space-time
variances~\cite{CSH95a,H96} of the emission function $S(x,K)$. 
These radius parameters depend in general on the azimuthal 
orientation of the  pair momentum ${\bf K}$ which we define 
with respect to the direction of the impact parameter $\vec{b}$,
  \begin{equation}
    \Phi = \angle(\vec{\bf K}_\perp,\vec{b})\, .
    \label{2.1}
  \end{equation}
They can be calculated as second derivatives
of the correlator $C({\bf K}, {\bf q})$ with respect to the
relative momentum components $i,j = o,s,l$ in the {\it osl}-system.
In what follows, we consider the emission function $S(x,K)$ to be 
given in the impact parameter fixed coordinate system. Then, to express
the HBT radii in terms of space-time variances, one has to 
rotate the coordinate system by the angle $\Phi$,
 \begin{mathletters}
   \label{2.2}
 \begin{eqnarray}
          &&({\cal D}_\Phi \vec{\beta}) =  \left( \begin{array}{c}
                 \beta_\perp \\ 0\\ \beta_l \end{array} \right)\, ,
           {\cal D}_\Phi \vec{\tilde{x}} =  \left( \begin{array}{c}
                 \tilde{x} \cos\Phi + \tilde{y} \sin\Phi  \\ 
                 -\tilde{x} \sin\Phi + \tilde{y} \cos\Phi \\
                 \tilde{z} \end{array} \right) \, , 
   \label{2.2a} \\
   && R_{ij}^2({\bf K}) = - \left. { \partial^2 C({\bf q},{\bf K})\over 
                           \partial q_i\, \partial q_j}
                           \right\vert_{ {\bf q} = 0}\, \nonumber \\
   && = \langle  \lbrack ({\cal D}_\Phi \tilde{x})_i 
                        - ({\cal D}_\Phi \beta)_i\tilde{t}\rbrack 
               \lbrack ({\cal D}_\Phi \tilde{x})_j 
                        - ({\cal D}_\Phi \beta)_j\tilde{t}\rbrack 
                        \rangle\, .
   \label{2.2b}
 \end{eqnarray}
 \end{mathletters}
Here, $\tilde{x}_\mu = x_\mu - \langle x_\mu\rangle$, and all 
coordinates $x$, $y$ and $z$ are given in the impact-parameter 
fixed system. The space-time variances specify the curvature of
the correlator at ${\bf q} = 0$ and coincide with the experimentally
determined half widths of $C({\bf K}, {\bf q})$ for Gaussian shapes 
only~\cite{WSH96,WH96}. Deviations from a Gaussian can be 
characterized by more
refined methods~\cite{WH96}. The present investigation
is restricted to correlators of sufficiently 
Gaussian shape and makes no attempt to quantify (possibly 
$\Phi$-dependent) non-Gaussian deviations. In the analysis of azimuthally 
symmetric HBT correlation radii, this Gaussian approximation 
has led to a qualitative and quantitative understanding 
of the $K_\perp$-dependence of correlation functions~\cite{H96}. This 
motivates us to adopt the same starting point for an azimuthally 
sensitive analysis. The six $\Phi$-dependent HBT 
radius parameters (\ref{2.2b}) read
  \begin{eqnarray}
    && R_s^2(K_\perp,\Phi,Y) = \langle \tilde{x}^2\rangle \sin^2\Phi
                  + \langle \tilde{y}^2\rangle \cos^2\Phi
                  - \langle \tilde{x}\tilde{y}\rangle 
                       \sin 2\Phi \, ,
                      \nonumber \\
    && R_o^2(K_\perp,\Phi,Y) = \langle \tilde{x}^2\rangle \cos^2\Phi
                  + \langle \tilde{y}^2\rangle \sin^2\Phi
                  + \beta_\perp^2 \langle \tilde{t}^2\rangle
                      \nonumber \\
    && \qquad \qquad - 2\beta_\perp 
                       \langle \tilde{t}\tilde{x} \rangle \cos\Phi 
                     - 2\beta_\perp 
                       \langle \tilde{t} \tilde{y} \rangle \sin\Phi 
                     + \langle \tilde{x}\tilde{y}\rangle 
                       \sin 2\Phi \, ,
                      \nonumber \\
    && R_{os}^2(K_\perp,\Phi,Y) = 
                  \langle \tilde{x}\tilde{y}\rangle \cos 2\Phi 
                  + \textstyle{1\over 2} \sin 2\Phi 
                  (\langle \tilde{y}^2\rangle - \langle \tilde{x}^2\rangle)
                  \nonumber \\
    && \qquad \qquad + \beta_\perp \langle \tilde{t}
                       \tilde{x}\rangle \sin\Phi
                     - \beta_\perp \langle \tilde{t}
                       \tilde{y}\rangle \cos\Phi \, ,
                      \nonumber \\
    && R_{l}^2(K_\perp,\Phi,Y) = 
                  \langle (\tilde{z} -\beta_l\tilde{t})^2 \rangle \, ,
                      \nonumber \\
    && R_{ol}^2(K_\perp,\Phi,Y) = 
                  \langle (\tilde{z} -\beta_l\tilde{t})
                     (\tilde{x}\cos\Phi + \tilde{y}\sin\Phi 
                     - \beta_\perp\tilde{t}) \rangle\, ,
                      \nonumber \\
    && R_{sl}^2(K_\perp,\Phi,Y) = 
                  \langle (\tilde{z} -\beta_l\tilde{t})
                     (\tilde{y}\cos\Phi - \tilde{x}\sin\Phi) 
                     \rangle\, .
    \label{2.3}
  \end{eqnarray}
These equations separate the {\it explicit} 
$\Phi$-dependence of the HBT-radii (which is a consequence 
of the changing direction of the pair momentum ${\bf K}$ with respect
to the reaction plane) from the {\it implicit} $\Phi$-dependence of the
spatio-temporal widths $\langle \tilde{x}_{\mu}\tilde{x}_{\nu}\rangle$
(which reflects a $\Phi$-dependent change of the shape of the
effective emission region). In general, both implicit and explicit 
$\Phi$-dependence will show up in the harmonic coefficients
  \begin{mathletters}
    \label{2.4}
  \begin{eqnarray}
    {R_{ij,m}^c}^2 &=& \textstyle{1\over 2\pi} 
           \int_{-\pi}^\pi R_{ij}^2\, \cos(m\Phi)\, d\Phi\, ,
    \label{2.4a} \\
    {R_{ij,m}^s}^2 &=& \textstyle{1\over 2\pi} 
           \int_{-\pi}^\pi R_{ij}^2\, \sin(m\Phi)\, d\Phi\, ,
    \label{2.4b}
  \end{eqnarray}
  \end{mathletters}
which determine the complete Gaussian parametrization (\ref{1.3}).

\begin{figure}[h]\epsfxsize=7cm 
\centerline{\epsfbox{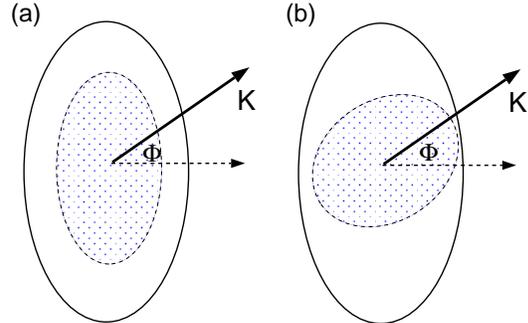}}
\caption{Schematic picture of a transverse cut through the collision
region. Shaded is the effective emission region contributing to the 
correlator for pair momentum ${K}$. Two different scenarios are shown: in
(a) the anisotropy is determined by the global geometry of the collision,
in (b) it is significantly influenced by the collective dynamics of the 
source. Both scenarios can be distinguished experimentally, see the
text.
}\label{fig1}
\end{figure}

The following discussion of the $\Phi$-dependent HBT radius
parameters (\ref{2.3}) is focussed mainly on the transverse parameters
$R_s^2$, $R_o^2$, $R_{os}^2$. Their harmonic coefficients depend 
on the $\Phi$-dependence of the transverse spatial widths
  \begin{equation}
    T_{ij}^\perp = \langle \tilde{x}_i \tilde{x}_j \rangle\, ,
    \label{2.5}
  \end{equation}
$i$, $j$ being components in the transverse plane. Any $\Phi$-dependence 
of $T^\perp$ is a consequence of non-trivial $x$-$\Phi$ correlations
and a fortiori of position-momentum 
correlations in the source. To illustrate this point, we have sketched 
in Fig.~\ref{fig1} two simplified scenarios. If there are no
$x$-$\Phi$-correlations, then the transverse shape of the effective 
emission region is $\Phi$-independent and reflects the global
geometry of the collision region. For non-trivial $x$-$\Phi$ 
correlations, this simple relation between the emission region and
the global geometry breaks down. In sections~\ref{sec3} and \ref{sec5},
we find observable combinations of first and second 
harmonic coefficients which are sensitive to this difference. 
More generally, we classify the possible 
$\Phi$-dependences of (\ref{2.5})
and discuss their implications for the harmonic analysis of 
HBT radius parameters. 
%
%
We conclude this section by shortly commenting on the
$\Phi$-dependence of $R_l^2$, $R_{ol}^2$ and $R_{sl}^2$.
The longitudinal radius parameter
$R_l^2$ shows no explicit $\Phi$-dependence and coincides 
formally with the expression for the azimuthally symmetric 
case~\cite{CSH95a}. 
The radii $R_{ol}^2$ and $R_{sl}^2$ contain explicit
$\Phi$-dependent terms proportional to 
$\langle \tilde{t}\tilde{x}\rangle$
or $\langle \tilde{t}\tilde{y}\rangle$. These characterize 
asymmetries of the particle emission probability 
around the point of highest emissivity and they vanish for 
models with Gaussian emission functions $S(x,k)$.
In the light of this, we consider in what follows 
all harmonic coefficients $m \geq 1$ of the HBT radii 
$R_l^2$, $R_{ol}^2$, $R_{sl}^2$ to be negligible,
  \begin{mathletters}
    \label{2.6}
  \begin{eqnarray}
    {R_{l,m}^c}^2 &=& {R_{l,m}^s}^2 \approx 0\, ,
    \label{2.6a} \\
    {R_{ol,m}^c}^2 &=& {R_{ol,m}^s}^2 \approx 0\, ,
    \label{2.6b} \\
    {R_{sl,m}^c}^2 &=& {R_{sl,m}^s}^2 \approx 0\, .
    \label{2.6c}
  \end{eqnarray}
  \end{mathletters}
This is a reasonable but model-dependent assumption.
For all models studied below, we have checked (\ref{2.6})
numerically, see section~\ref{sec42}.

\section{Azimuthal analysis of the mid-rapidity region}
\label{sec3}

In this section, we discuss the azimuthal analysis of 1- and
2-particle spectra for symmetric collision systems like 
Pb-Pb or Au-Au at mid-rapidity. The important simplification 
at mid-rapidity is that all observables are 
invariant under $\Phi \to \Phi + \pi$, a $180^o$-rotation
in the transverse plane. All odd harmonic coefficients vanish. 
This is different for the fragmentation region
or for non-symmetric collision systems where the only remaining
symmetry is that with respect to the reaction plane. 
The arising complications are discussed in section~\ref{sec5}.
Here, we first discuss
a scenario, for which the space-time variances (\ref{2.5}) do 
not depend on the azimuthal direction of ${\bf K}_\perp$.
Then, we turn to the discussion of an arbitrary $\Phi$-dependence
of $T^\perp$.
%
\subsection{The elliptic approximation}
\label{sec31}

We start by considering an elliptic approximation of the transverse 
spatial widths $T_{ij}^\perp$. This toy example will be useful in the 
sequel for comparisons with the general case. It provides a simple 
picture for the consequences of a purely geometrical scenario, and
it can account for some of the main features of the harmonic 
coefficients, calculated in the model study in section~\ref{sec4}.

In the elliptic approximation, we characterize the tensor $T^\perp$
by its principal axes (eigenvectors) $\vec{g}_1$, $\vec{g}_2$, and 
the corresponding eigenvalues $g_1$ and $g_2$. If the orientation of
the reaction plane and $\vec{g}_1$ differs by an angle $\varphi$,
then $T_{ij}^\perp$ takes a simple form in the impact-parameter 
fixed system, 
  \begin{mathletters}
    \label{3.1}
  \begin{eqnarray}
   T^\perp &=& \bar{g} \left( \begin{array}{cc}
                1+\alpha_T\cos 2\varphi & -\alpha_T\sin 2\varphi \\
                -\alpha_T\sin 2\varphi  & 1-\alpha_T\cos 2\varphi 
                 \end{array} \right)\, ,
   \label{3.1a}\\
   \bar{g} &=& \textstyle{1\over 2}(g_1+g_2)\, ,\, \qquad 
   \alpha_T = { {g_1-g_2}\over {g_1+g_2} } \, .
    \label{3.1b}
  \end{eqnarray}
  \end{mathletters}
This is somewhat analogous to the expression of the transverse 
sphericity tensor in \cite{O92,O93}. It provides a convenient 
parametrization of the three independent expectation values
$\langle \tilde{x}_i \tilde{x}_j \rangle$ in terms of the average 
size $\bar{g}$ of the homogeneity region, its transverse spatial
anisotropy $\alpha_T$ and the orientation $\varphi$ of its principal
axis $\vec{g}_1$ with respect to the reaction plane. 
In case of an azimuthally symmetric collision with vanishing 
impact parameter, $\varphi = 0$, the cross term vanishes,
$\langle \tilde{x}_s \tilde{x}_o \rangle = 0$, and  
the spatial asymmetry $\alpha_T$ in (\ref{3.1a}), (\ref{3.1b}) 
describes the difference between the spatial widths in the 
{\it out}- and {\it side}-directions. In terms of these parameters,
the `transverse' HBT radii $R_o$, $R_s$ and $R_{os}$ read 
  \begin{mathletters}
    \label{3.2}
  \begin{eqnarray}
    R_s^2(K_\perp,\Phi,Y) &=& \bar{g} 
                 {\left({ 1-\alpha_T\cos 2(\varphi + \Phi) }\right)} \, ,
                  \label{3.2a} \\
    R_o^2(K_\perp,\Phi,Y) &=& \bar{g} 
                 {\left({ 1+\alpha_T\cos 2(\varphi + \Phi) }\right)}
                 \nonumber \\
                  && + \beta_\perp^2 \langle \tilde{t}^2\rangle + C_o\, ,
                  \label{3.2b} \\
    R_{os}^2(K_\perp,\Phi,Y) &=& 
                  - \bar{g} \alpha_T\sin 2(\varphi + \Phi) + C_{os}\, ,
                  \label{3.2c} \\
             C_o  &=& - 2\beta_\perp 
                       \langle \tilde{t}\tilde{x} \rangle \cos\Phi 
                     - 2\beta_\perp 
                       \langle \tilde{t} \tilde{y} \rangle \sin\Phi \, ,
                  \label{3.2d} \\
            C_{os}  &=& + \beta_\perp \langle \tilde{t}
                       \tilde{x}\rangle \sin\Phi
                     - \beta_\perp \langle \tilde{t}
                       \tilde{y}\rangle \cos\Phi \, .
    \label{3.2e}
  \end{eqnarray}
  \end{mathletters}
The two correction terms $C_o$, $C_{os}$ 
contain widths $\langle \tilde{t} \tilde{x}_i \rangle$
which are linear in $\tilde{t}$ and measure asymmetries of the
source around the point of highest emissivity. At mid-rapidity,
all observables are invariant under $\Phi \to \Phi + \pi$,
and $\langle \tilde{x}\tilde{t} \rangle(\Phi + \pi)
= - \langle \tilde{x}\tilde{t}\rangle(\Phi)$.
The correction terms hence do {\it not}
contribute to the first harmonic coefficients ${R_{ij,1}}^2$, but
to the second ones. Moreover, they vanish for emission functions 
$S(x,K)$ which are Gaussian in the spatial components.
Their contribution to $R_{o}^2$, $R_{os}^2$ is neglected in the
remainder of this section, and the validity of this approximation
is checked in the numerical model study of section \ref{sec4}.
 
The main assumption of the elliptic approximation which we 
avoid in section~\ref{sec32}, is that the
parameters $\bar{g}$ and $\alpha_T$
do not depend on the azimuthal orientation of ${K}_\perp$.
The homogeneity region is determined entirely by the geometry, 
cf. Fig.~\ref{fig1}a. The eigenvectors $\vec{g}_1$, $\vec{g}_2$ 
lie parallel and orthogonal to the reaction plane and the 
angle $\varphi$ should vanish; it hence accounts only for the 
statistical uncertainty in determining the reaction plane, see 
section~\ref{sec6}. Comparing the expressions (\ref{3.2}) with the 
Fourier expansion (\ref{1.3}) of the HBT radius parameters, we find
for $\varphi = 0$
  \begin{mathletters}
    \label{3.3}
  \begin{eqnarray}
    {R_{o,0}}^2 &=& \bar{g} 
           + \beta_\perp^2 \langle \tilde{t}^2\rangle\, , \qquad
    {R_{s,0}}^2 = \bar{g}\, ,
    \label{3.3a} \\
    {R_{o,2}^c}^2 &=& - {R_{s,2}^c}^2 = - {R_{os,2}^s}^2
                   = \textstyle{1\over 2}\alpha_T \bar{g}\, . 
    \label{3.3b}
  \end{eqnarray}
  \end{mathletters}
The other second and higher order Fourier coefficients vanish.
According to (\ref{3.3}), the ansatz (\ref{1.3}) for the correlator
contains redundant information. In the present
case, neither the shape of $T^\perp$ nor the
size of the homogeneity region depend on $\Phi$ while the
anisotropy terms of the HBT radius parameters are sufficiently
many to encode for such a non-trivial $\Phi$-dependence. 

\subsection{The general case}
\label{sec32}

To discuss an arbitrary $\Phi$-dependence of the transverse 
spatial widths, we start from a complete Fourier expansion of 
$T^\perp_{ij} = \langle \tilde{x}_i\tilde{x}_j\rangle$,
respectively of its linear combinations
  \begin{eqnarray}
    A &=& \textstyle{ {\langle \tilde{x}^2\rangle 
                        + \langle \tilde{y}^2\rangle}\over 2}
       = \sum_{n=0}^\infty \left( A_n\cos n\Phi 
                      + A'_n\sin n\Phi \right)\, ,
    \nonumber \\
    B &=& \textstyle{ {\langle \tilde{x}^2\rangle 
                        - \langle \tilde{y}^2\rangle}\over 2} 
       = \sum_{n=0}^\infty \left( B_n\cos n\Phi 
                      + B'_n\sin n\Phi \right)\, ,
    \nonumber \\
    C &=& \langle\tilde{x}\tilde{y}\rangle 
       = \sum_{n=0}^\infty \left( C'_n\cos n\Phi 
                      + C_n\sin n\Phi \right)\, .
    \label{3.4}
  \end{eqnarray}
Here, the coefficients $A_n$, $A'_n$, etc.
are functions of the longitudinal pair rapidity 
$Y$ and the transverse pair momentum $\vert {\bf K}_\perp\vert$, only.
The symmetry of the collision region with respect to the reaction 
plane implies that the terms $\langle \tilde{x}^2\rangle$,
$\langle \tilde{y}^2\rangle$ are invariant under $\Phi \to -\Phi$,
while the term $\langle\tilde{x}\tilde{y}\rangle$ changes its sign.
Calculating the coefficients in (\ref{3.4}) via Fourier transform,
it is easy to check that the primed ones vanish,
  \begin{equation}
    A'_n = B'_n = C'_n = 0 \, .
  \label{3.5}
  \end{equation}
The general $\Phi$-dependent HBT radius parameters (\ref{2.3})
read 
  \begin{eqnarray}
    R_s^2 &=& A - B\cos 2\Phi - C\sin 2\Phi\, ,
    \nonumber \\
    R_o^2 &=& A + B\cos 2\Phi + C\sin 2\Phi 
           + \beta_\perp^2\, \langle\tilde{t}^2\rangle \, ,
    \nonumber \\
    R_{os}^2 &=& - B\sin 2\Phi + C\cos 2\Phi\, ,
    \label{3.6}
  \end{eqnarray}
where we have dropped the small correction terms $C_o$, $C_{os}$.
From these HBT radius parameters (\ref{3.6}) one can
calculate the harmonic coefficients (\ref{2.4}). 
Especially, we obtain the zeroth harmonics
  \begin{eqnarray}
    R_{s,0}^2 &=& A_0 - \textstyle{1\over 2}B_2 
                      - \textstyle{1\over 2}C_2\, ,
    \nonumber \\
    R_{o,0}^2 &=& A_0 + \textstyle{1\over 2}B_2 
                      + \textstyle{1\over 2}C_2
                      + \beta_\perp^2\, \langle\tilde{t}^2\rangle \, ,
    \nonumber \\
    R_{os,0}^2 &=& 0\, ,
    \label{3.7}
   \end{eqnarray}
and the second harmonic coefficients 
   \begin{eqnarray}
     {R_{s,2}^c}^2 &=& - \textstyle{1\over 2}B_0
                       + \textstyle{1\over 2}A_2 
                       - \textstyle{1\over 4}B_4 
                       - \textstyle{1\over 4}C_4 \, ,
     \nonumber \\
     {R_{o,2}^c}^2 &=& \textstyle{1\over 2}B_0
                       + \textstyle{1\over 2}A_2 
                       + \textstyle{1\over 4}B_4 
                       + \textstyle{1\over 4}C_4\, ,
     \nonumber \\
     {R_{os,2}^s}^2 &=& - \textstyle{1\over 2}B_0
                       + \textstyle{1\over 4}B_4 
                       + \textstyle{1\over 4}C_4\, ,
     \nonumber \\               
     {R_{s,2}^s}^2 &=& {R_{o,2}^s}^2 = {R_{os,2}^c}^2 = 0\, .  
    \label{3.8}
  \end{eqnarray}
For the case of $\Phi$-independent spatial widths $T^\perp$, 
only the terms with index $0$ survive on the right hand side, 
and (\ref{3.8}) coincides with (\ref{3.3b}) for
$A_0 = \bar{g}$ and $B_0 = \alpha_T\bar{g}$. If deviations from 
(\ref{3.3b}) are observed, this is an unambiguous sign for 
source gradients leading to a $\Phi$-dependence of $T_{ij}^\perp$.
From the absence of such deviations, however, one cannot conclude
that there are no source gradients. In the following model study
we shall find that even in the presence of sizeable source 
gradients, such deviations can be small. Then, the use of 
equations (\ref{3.3b}) resides in reducing the number of fit
parameters in an azimuthal HBT-analysis, see section~\ref{sec6}.

Let us finally anticipate that for the models studied below,
the fourth harmonic coefficients $A$, $B$ and $C$ are
negligible. It follows from (\ref{3.8}) that then the deviations 
of (\ref{3.3b}) are essentially determined by the term 
$\textstyle{1\over 2} A_2$ only, i.e.,  
  \begin{equation}
     {R_{o,2}^c}^2 + 2\, {R_{os,2}^s}^2 \approx {R_{s,2}^c}^2\, .
     \label{3.9}
  \end{equation}
%
\section{A model calculation}
\label{sec4}

We introduce now a simple 
hydrodynamical model for the emission function of a heavy ion collision
which includes anisotropy effects. For this model, we calculate both
the one- and two-particle spectra, illustrating the main points of
the above model-independent discussion. 

In the central rapidity region of a peripheral collisions,
the initial distribution of the highly excited nuclear matter 
is given by the intersection of the nuclear
spheres. The largest pressure gradient developping from such initial 
conditions is expected to be aligned with the impact parameter 
$\vec{b}$, cf. ~\cite{O92,F96}. To mimic this scenario, we consider 
the class of model emission functions
 \begin{equation}
   S(x,K) = \tau_0 m_\perp \cosh(\eta-y)\, 
            \exp\left(\textstyle{-K^\mu u_\mu(x)\over T}\right)\, H(x)\, ,
   \label{4.1}
 \end{equation}
whose azimuthally symmetric versions have been discussed extensively
in the literature. For a review, cf.~\cite{H96}.
These models assume the emission of particles
from a thermalized system with collective four-velocity $u_\mu$, 
confined in a space-time volume determined by $H(x)$. The factor 
$P\cdot n(x) = \tau_0 m_\perp \cosh(\eta-y)$ specifies a simple
hyperbolic freeze-out hypersurface. We introduce an azimuthal 
anisotropy in the source both via an elliptic shape of the geometrical 
emission region,
 \begin{equation}
   H(x) = \exp \left[- {(\tau-\tau_0)^2 \over 2(\Delta \tau)^2}
                       - {{(\eta- \eta_0)}^2 \over 2 (\Delta \eta)^2}
                       - {x^2 \over 2 \rho_x^2} - {y^2 \over 2 \rho_y^2}
           \right] ,
   \label{4.2}
 \end{equation}
and via an azimuthally asymmetric flow pattern $u_\mu(x)$ which
is properly normalized, $u_\mu u^\mu = 1$,
  \begin{eqnarray}
    u_\mu(x) &=& (u_l\cosh\eta, u_x, u_y,
                       u_l\sinh\eta )\, ,
                       \nonumber \\
    u_x &=& \textstyle{x\over \lambda_x}\, ,\, \, 
    u_y = \textstyle{y\over \lambda_y} \, ,\, \, 
    u_l = \sqrt{1+u_x^2 + u_y^2}\, .
    \label{4.3}
  \end{eqnarray}
In the longitudinal direction, we choose a boost invariant
flow pattern satisfying Bjorken scaling~\cite{B83}, i.e., 
the main energy flow is along the beam axis.
Instead of the variables $\rho_x$, $\rho_y$, $\lambda_x$, $\lambda_y$,
we use in what follows the transverse size $R$ and its spatial 
anisotropy $\epsilon_s$, as well as the transverse flow strength
$\eta_f$ and the corresponding flow anisotropy $\epsilon_f$,
  \begin{mathletters}
    \label{4.4}
  \begin{eqnarray}
    \rho_x &=& R \sqrt{1-\epsilon_s}\, ,
    \qquad
    \rho_y = R \sqrt{1+\epsilon_s}\, ,
    \label{4.4a} \\
    u_x &=& \eta_f \sqrt{1+\epsilon_f}\textstyle{x\over R}\, ,
    \qquad
    u_y = \eta_f \sqrt{1-\epsilon_f}\textstyle{y\over R}\, .
    \label{4.4b}
  \end{eqnarray}
  \end{mathletters}
We have chosen the principal axes of the transverse space-time 
distribution aligned with those of the azimuthal momentum distribution, 
since the dynamical evolution of the collision region cannot break 
the reflection symmetry of the system with respect to the reaction
plane. The spatial anisotropy $\epsilon_s$ takes values in the
range $-1 < \epsilon_s < 1$, i.e., for $\epsilon > 0$, the
collision region is longer in the direction perpendicular to $\vec{b}$,
$\rho_x < \rho_y$. For the flow anisotropy $\epsilon_f$, we allow for
$-1 < \epsilon_f < 1$; the major flow component lies in the reaction
plane if $\epsilon_f$ is positive. All numerical calculations are done 
for the input parameters $T = 150$ MeV, $m = m_\pi = 139$ MeV, 
$\tau_0 = 5$ fm/c, $\Delta\tau = 1$ fm/c, $\Delta\eta = 1.22$ and
$R = 5$ fm. We study the dependence of the 1- and 2-particle spectra
on the size of the transverse flow, and the spatial $\epsilon_s$ and
dynamical $\epsilon_f$ anisotropies.

\subsection{Harmonic analysis of azimuthal particle distributions}
\label{sec41}
The harmonic coefficients $v_n$ of the triple-differential 
one-particle spectrum (\ref{1.1}) are given in terms of the 
Fourier transforms
  \begin{equation}
    \left(\begin{array}{c} a_n \\ b_n \end{array} \right)
     = { {\int_0^{2\pi} E\textstyle{dN\over d^3p} 
    \left({ \begin{array}{c} \cos(n\phi) \\ \sin(n\phi) \end{array} }\right)
    \, d\phi} \over
    {\int_0^{2\pi} E\textstyle{dN\over d^3p} d\phi} }\, ,
    \label{4.5a}
  \end{equation}
  \begin{equation}
    a_n = v_n\, \cos(n\psi_R^{(n)})\, ,\qquad
    b_n = v_n\, \sin(n\psi_R^{(n)})\, .
    \label{4.5b}
  \end{equation}
The zeroth harmonic $v_0$ denotes the azimuthally averaged double
differential particle distribution, all other anisotropy parameters 
$v_n$ characterize azimuthal asymmetries in the momentum distribution.

In the absence of transverse flow, $\eta_f = 0$, the model (\ref{4.1}) 
does not contain source gradients in the transverse plane. Irrespective
of whether the transverse geometry of the source is radially symmetric 
($\epsilon_s = 0$) or not, particle emission is  isotropic in
the transverse plane. All harmonic coefficients $v_n$,
$n \geq 1$, vanish.

In the presence of transverse flow, the reflection 
symmetry of the emission function (\ref{4.1}) in the direction of
the impact parameter implies a 
$\phi \to \phi + \pi$ symmetry of the particle spectra. 
All odd harmonic coefficients $v_n$ vanish. The lowest non-vanishing
anisotropy parameter is $v_2$. This parameter is positive if
the spectrum $d^3N/(p_t\, dp_t\, dy\, d\phi)$ shows a maximum in the
reaction plane, it is negative for the opposite case. Higher
fourth (6th, etc.) harmonic coefficients are found to be much
smaller. They are not discussed further.

\begin{figure}[h]\epsfxsize=8cm
\centerline{\epsfbox{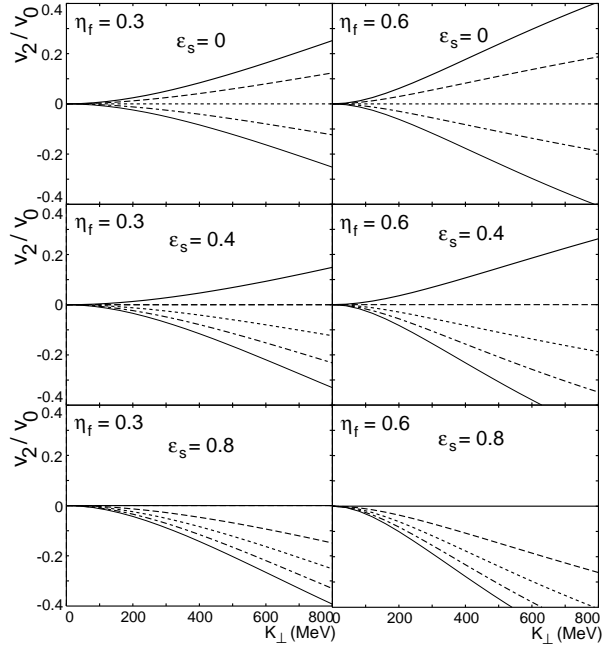}}
\caption{The normalized second harmonic coefficients $v_2/v_0$ of
the single particle spectrum for the model (\protect{\ref{4.1}}). 
The different plots show results for scenarios with different 
transverse flow strengths $\eta_f$ and geometrical anisotropies
$\epsilon_s$. The lines denote different
transverse flow anisotropies: $\epsilon_f = 0.8$ (solid line),
$\epsilon_f = 0.4$ (dashed line), $\epsilon_f = 0$ (dotted line),
$\epsilon_f = -0.4$ (dash-dotted line) and $\epsilon_f = - 0.8$ (thin 
solid line). Positive values of $\epsilon_f$ correspond to the major
flow direction lying in the reaction plane. 
}\label{fig2}
\end{figure}

The transverse pair momentum dependence of the corresponding normalized 
quantity $v_2/v_0$ is depicted in Fig.~\ref{fig2} for different physical 
scenarios. Irrespective of the model parameters $\eta_f$, $\epsilon_s$ and
$\epsilon_f$, the coefficient $v_2/v_0$ vanishes at $K_\perp = 0$
and is growing monotonously with the transverse pair momentum.
This is a direct consequence of the Lorentz-invariant Boltzmann term
in (\ref{4.1}) which encodes the assumption of local thermal
equilibrium, 
  \begin{equation}
    K^{\mu}u_{\mu} = \textstyle{{m_\perp}\over T}\cosh(\eta-Y)\, u_l
       - \textstyle{K_x\over T}u_x - \textstyle{K_y\over T}u_y\, .
    \label{4.6}
  \end{equation}
The terms which couple the 
flow components $u_x$, $u_y$ linear to the transverse
momentum components are the only ones in the model
emission function (\ref{4.1}) which can introduce a $\Phi$-dependence.
In the limit $K_\perp \to 0$, these $\Phi$-dependent terms  
vanish, the emission probabilities in different azimuthal 
directions become equal, and 
  \begin{equation}
    \lim_{K_\perp\to 0} v_n(K_\perp) = 0\, \qquad
    \hbox{for all $n\geq 1$.}
    \label{4.7}
  \end{equation}
More explicit expressions for the parameters $v_n$ of this model
can be obtained in a saddle-point approximation, details of
which are presented in Appendix~\ref{appa}. In this approximation,
we find that $v_2 \propto K_\perp^2$ for small values of $K_\perp$.
For small transverse flow, the leading dependence on the
anisotropy parameters $\epsilon_s$ and $\epsilon_f$ is given by
  \begin{equation}
    v_2 \propto \left( {\lambda_y^2\over \rho_y^2}
                     - {\lambda_x^2\over \rho_x^2} \right)
        \propto { {2(\epsilon_f - \epsilon_s)}\over 
                   (1-\epsilon_f^2)\, (1-\epsilon_s^2) }\, .
    \label{4.8}
  \end{equation}
%
%
Equation (\ref{4.8}) describes correctly the main features of 
the numerical results in Fig.~\ref{fig2}. The coefficient
$v_2(K_\perp)$ vanishes for $\epsilon_f = \epsilon_s$, 
and its sign coincides with 
that of $\epsilon_f-\epsilon_s$. In the case of an azimuthally 
symmetric particle emission region, $\epsilon_s = 0$, this behaviour
is entirely due to the transverse flow anisotropy $\epsilon_f$. For
positive values of $\epsilon_f$, the major transverse flow component
lies in the reaction plane and $v_2$ is positive. Negative values of
$\epsilon_f$ mimic a squeeze-out scenario where the major flow component
lies orthogonal to the reaction plane. This explains the 
$\epsilon_f$-dependence of $v_2$, shown in Fig.~\ref{fig2}.
On the other hand, if we choose
for a given flow pattern ($\eta_f$, $\epsilon_f$ fixed) more and more 
elongated transverse geometries, then $v_2$ decreases. The reason is
that increasing 
$\epsilon_s$ results in more emission points with a large $u_y$ and 
small $u_x$ flow component. It thus mimics a larger squeeze-out 
component of the transverse flow orthogonal to the reaction plane.
In the simple model discussed here, increasing $\epsilon_s$ and 
decreasing $\epsilon_f$ hence affects the azimuthal particle distributions 
similarly. Spatial and dynamical information cannot be disentangled 
completely on the basis of single-particle spectra. For azimuthally 
symmetric scenarios, this is well-known~\cite{SH94}.
%
\subsection{Harmonic analysis of HBT radius parameters}
\label{sec42}

Here, we check first that the $\Phi$-dependence 
of the radii $R_l^2$, $R_{ol}^2$, and $R_{sl}^2$ is negligible.
At mid-rapidity, $Y = \beta_l = 0$, the 
emission function (\ref{4.1}) is symmetric with respect to $z\to - z$,
and the HBT radius parameters $R_{ol}^2$ and $R_{sl}^2$ vanish,
  \begin{eqnarray}
    && R_{ol}^2\, ,R_{sl}^2 \propto
    \langle (\tilde{z} -\beta_l\tilde{t})\rangle\, , 
    \nonumber \\
    && \langle (\tilde{z} -\beta_l\tilde{t})\rangle\vert_{Y=0} = 0\,
    \qquad \hbox{for all ${\bf K}_\perp$.}
    \label{4.9}
  \end{eqnarray}
%
\begin{figure}[h]\epsfxsize=7cm 
\centerline{\epsfbox{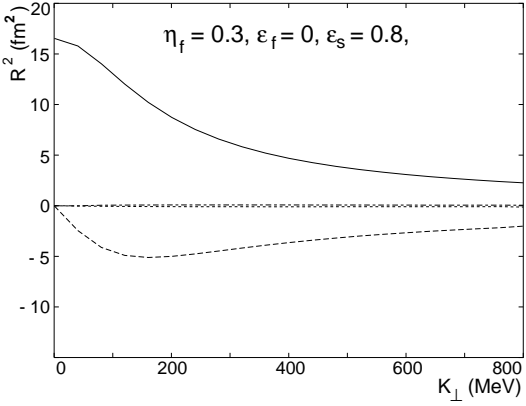}}
\caption{The non-vanishing zeroth and second harmonic coefficients
of the non-transverse HBT radius parameters, ${R_{l,0}}^2$ (solid line), 
${R_{ol,0}}^2$ (dashed
line), ${R_{ol,2}^c}^2$ (dotted line) and ${R_{sl,2}^s}^2$ 
(dash-dotted line) for forward rapidity $Y = 1$.
The second harmonic coefficients
${R_{ol,2}^c}^2$ and ${R_{sl,2}^s}^2$ are proportional to 
$\langle\tilde{x}\rangle$ and $\langle\tilde{y}\rangle$,
which are generically small correction terms, see the discussion
following \protect{(\ref{4.9})}.
}\label{fig3}
\end{figure}
%
To study their ${\bf K}_\perp$-dependence, we hence choose 
the forward rapidity $Y = 1$. Numerical results are presented in 
Fig.~\ref{fig3}. The zeroth harmonic coefficients show the  
$K_\perp$-dependence expected from azimuthally symmetric model
studies~\cite{WSH96}. Especially, the longitudinal radius 
parameter $R_{l,0}^2$ has a steep $K_\perp$-slope which reflects 
the strong longitudinal expansion of the source. Also, the 
out-longitudinal cross term $R_{ol,0}^2$ shows the typical 
$K_\perp$-dependence known from studies of azimuthally symmetric
models. It takes significant non-zero values and vanishes at
$K_\perp = 0$ where $R_{ol}^2 = R_{sl}^2$. The parameter
$R_{sl,0}^2$ vanishes for all $K_\perp$. For an azimuthally
symmetric emission region, this would be a consequence of the
$q_s \to -q_s$ reflection symmetry of the correlator. Contributions
breaking this symmetry introduce automatically a $\Phi$-dependence
and hence do not show up in the zeroth harmonics.
Higher harmonics are found to be very small. 
$R_l^2$ shows no $\Phi$-dependence, the only non-vanishing 
second harmonics are ${R_{sl,2}^s}^2$ and ${R_{ol,2}^c}^2$.
From Fig.~\ref{fig3} we conclude that these higher harmonics
are negligible. This illustrates the arguments leading to
(\ref{2.6}). It suggests to restrict an azimuthally sensitive HBT 
analysis to the `transverse' HBT radius parameters $R_o^2$, $R_s^2$ 
and $R_{os}^2$. 
%
\begin{figure}[h]\epsfxsize=7cm 
\centerline{\epsfbox{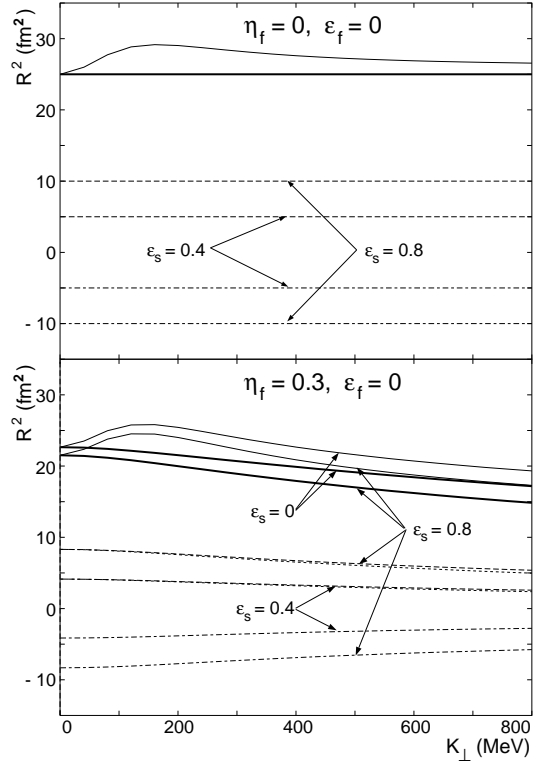}}
\caption{Zeroth and second harmonic coefficients for the transverse
HBT radius parameters of the model (\protect{\ref{4.1}}) without 
($\eta_f = 0$) and with ($\eta_f = 0.3$) transverse flow, and 
for different sizes of the spatial anisotropy $\epsilon_s$. Thin and
thick solid lines denote the zeroth coefficients $R_{o,0}^2$ and 
$R_{s,0}^2$ respectively. The plot shows all nonvanishing second 
harmonic coefficients, ${R_{o,2}^c}^2$ (dash-dotted line), 
${R_{s,2}^c}^2$ (dashed line) and ${R_{os,2}^s}^2$ (dotted line).
For $\eta_f = 0$, ${R_{s,2}^c}^2 = {R_{os,2}^s}^2$, see (\ref{3.3b}).
}\label{fig4}
\end{figure}
%
For the transverse HBT radii of the model (\ref{4.1}), suggestive 
analytical expressions
can be obtained in the approximation $u_l \approx 1
+ \textstyle{1\over 2}u_x^2 + \textstyle{1\over 2}u_y^2$. 
Deferring all technical details to Appendix~\ref{appa}, we merely
state that in this approximation, the $\Phi$-dependence of
$T^\perp$ is lost. The zeroth and second harmonic coefficients of
all transverse HBT radii can be written in terms of an average size 
$\bar{g}$ and a spatial anisotropy $\alpha_T$, introduced in
section~\ref{sec3}, Eq. (\ref{3.3}),
  \begin{mathletters}
    \label{4.10}
  \begin{eqnarray}
    \bar{g}\alpha_T &=& { - R^2 \left(
      {\epsilon_s + \epsilon_f\, \textstyle{m_\perp\over T}\eta_f^2\, 
                   (1-\epsilon_s^2)}\right)
               \over
               {1 + 2\textstyle{m_\perp
                     (1-\epsilon_s\epsilon_f) \over T}\eta_f^2 
                  + \textstyle{m_\perp^2(1-\epsilon_s^2)\, (1-\epsilon_f^2)
                    \over T^2}\eta_f^4} }\, ,
    \label{4.10a} \\
  \bar{g} &=& { R^2 \left(
       {1 + \textstyle{m_\perp\over T}\eta_f^2\, (1-\epsilon_s^2)}\right)
               \over
               {1 + 2\textstyle{m_\perp
                     (1-\epsilon_s\epsilon_f) \over T}\eta_f^2 
                  + \textstyle{m_\perp^2(1-\epsilon_s^2)\, (1-\epsilon_f^2)
                    \over T^2}\eta_f^4} }\, .
    \label{4.10b}
  \end{eqnarray}
  \end{mathletters}
In the limit of vanishing transverse flow, $\eta_f \to 0$, 
these expressions become exact. The value of $\bar{g}$
which determines the zeroth harmonic of $R_s^2$ is just given by the
transverse geometrical radius $R^2$. The anisotropy $\bar{g}\alpha_T$
which specifies the second harmonic coefficients, is proportional to
$\epsilon_s$. This is the case shown in Fig.~\ref{fig4}. For
$\eta_f = 0$, the model emission function has no intrinsic  
position-momentum correlations, and the different harmonic coefficients 
satisfy the relations (\ref{3.3}).
The components ${R_{s,2}^c}^2$ and ${R_{os,2}^s}^2$
coincide and they differ from ${R_{o,2}^c}^2$ by an overall sign
only.

In the presence of realistic transverse flow strengths $\eta_f$,
the approximation $u_l \approx 1
+ \textstyle{1\over 2}u_x^2 + \textstyle{1\over 2}u_y^2$ looses
its validity. Numerical calculations are needed
to make precise quantitative statements, but the 
expressions (\ref{4.10}) still describe essential qualitative
features. In the limiting case of an azimuthally symmetric collision
region with a finite transverse flow $\eta_f$, the anisotropy parameter
$\alpha_T$ vanishes and $\bar{g}$ goes to the well-known~\cite{CSH95a} 
lowest order expression for the side radius parameter,
  \begin{equation}
    {R_{s}}^2 = \bar{g} = {R^2\over \left(
       {1 + \textstyle{m_\perp\over T}\eta_f^2 }\right)}\, \qquad
    \hbox{for $\epsilon_s=\epsilon_f=0$.}
    \label{4.11}
  \end{equation}
This describes the leading $m_\perp$-dependence of ${R_{s,0}}^2$ 
as a function of $\eta_f$: the slope of the side radius parameter
is indicative of the transverse flow $\eta_f$.

In the absence of dynamical anisotropies ($\epsilon_f = 0$)
the second harmonics ${R_{o,2}^c}^2$, ${R_{s,2}^c}^2$ 
and ${R_{os,2}^s}^2$ given essentially by 
$\textstyle{1\over 2}\bar{g}\alpha_T$, are sensitive to the 
strength of the geometrical anisotropy $\epsilon_s$. 
For small values of $\eta_f^2$, the leading $\epsilon_s$-dependence
of (\ref{4.10a}) is linear and this is consistent with
the scaling of the dashed and dash-dotted ($\epsilon_f = 0$) lines in 
Fig.~\ref{fig4}. The $K_\perp$-slopes depicted in Fig.~\ref{fig4} are 
qualitatively explained by (\ref{4.10a}). Also, we have investigated
numerically the $\epsilon_f$-dependence of the HBT radius parameters
for the model of section~\ref{sec41}. Here, we merely state that
the main qualitative features of the numerical results 
can be understood in terms of the analytical approximations (\ref{4.10a}),
(\ref{4.10b}).

\begin{figure}[h]\epsfxsize=7cm 
\centerline{\epsfbox{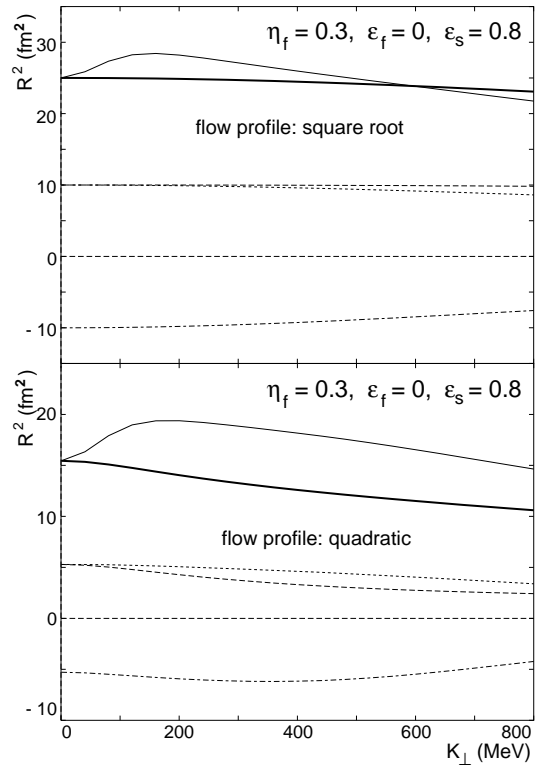}}
\caption{Same as Fig.~\protect{\ref{fig4}}, but for a square root
and a quadratic transverse flow profile, for which a Gaussian 
approximation of the emission function is not $\Phi$-independent.
In these cases, the difference ${R_{s,2}^c}^2 - {R_{os,2}^s}^2$
is more significant, but ${R_{o,2}^c}^2$ $+ 2{R_{os,2}^s}^2$ 
$\approx$ ${R_{s,2}^c}^2$ is still valid, see Eq. 
(\protect{\ref{3.9}}) and text below.
}\label{fig5}
\end{figure}

It is a remarkable feature of the model (\ref{4.1}) that the transverse
spatial widths $T_{ij}^\perp$ calculated in a saddle-point approximation
do not show any $\Phi$-dependence. This can be traced back to the 
the $\Phi$-dependent terms in (\ref{4.6}) being linear 
in $x$ and $y$, rather than e.g. quadratic. The slight
difference ${R_{s,2}^c}^2 - {R_{os,2}^s}^2$ found in the numerical
calculation of Fig.~\ref{fig4} for $\eta_f = 0.3$ stems from
the error made in the Gaussian approximation. To avoid 
drawing conclusions on the basis of a very model-dependent
feature like this almost complete cancelation of $\Phi$-dependent
contributions for linear flow profiles, we have investigated 
the following two non-linear flow profiles as well,
  \begin{mathletters}
    \label{4.12}
  \begin{eqnarray}
    u_i &=& \eta_f \sqrt{1\pm \epsilon_f}\, 
    \textstyle{x_i\over \sqrt{R \vert x_i\vert}}
    \, ,\qquad \hbox{[square root profile],}
    \label{4.12a} \\
    u_i &=& \eta_f \sqrt{1\pm \epsilon_f}\, 
    \textstyle{x_i \vert x_i\vert\over R^2}
    \, ,\qquad \hbox{[quadratic profile].}
    \label{4.12b}
  \end{eqnarray}
  \end{mathletters}
Here the index $i$ runs over $x$, $y$. Results for vanishing flow anisotropy
$\epsilon_f$ are shown in Fig.~\ref{fig5}. In the limit $K_\perp \to 0$,
the dependence of $T^\perp$ on the azimuthal direction of $K_\perp$
has to vanish and hence, all scenarios depicted in Figs.~\ref{fig4}
and \ref{fig5}, confirm the purely geometrical relation
(\ref{3.3b}), ${R_{o,2}^c}^2 =$ $- {R_{s,2}^c}^2 =$
$- {R_{os,2}^s}^2$. For increasing values of $K_\perp$, we find
deviations from this relation which are clearly more 
significant for the non-linear flow profiles. These however are
very well accounted for by the modified relation (\ref{3.9}) 
amongst the second harmonic coefficients, ${R_{o,2}^c}^2$ 
$+ 2\,{R_{os,2}^s}^2$ $= {R_{s,2}^c}^2$. The deviations from a 
purely geometrical scenario seen in Fig.~\ref{fig5}
are hence not due to the correction terms $C_o$, $C_{os}$, which are
generically negligible, but to the leading $\Phi$-dependence 
of $T^\perp$ which enters (\ref{3.7}) via the term $A_2$.
%
%
For the sake of completeness, the zeroth harmonic coefficients 
are presented in Fig.~\ref{fig5} as well. 
Details of their $K_\perp$-dependence 
can be understood by investigating the dependence of
$\langle \tilde{x}^2\rangle$ and $\langle \tilde{y}^2\rangle$ 
on the  pair momentum~\cite{comment}. 

Let us sum up the results of our model study: All second harmonic 
coefficients of non-transverse HBT radius parameters are negligible. 
Amongst the transverse HBT radii,
three second harmonic coefficients are non-negligible, 
namely ${R_{o,2}^c}^2$, ${R_{s,2}^c}^2$, and ${R_{os,2}^s}^2$.
Their leading contribution is determined by one parameter only,
see (\ref{3.3b}). The leading deviation from (\ref{3.3b}) satisfies
(\ref{3.9}) and is indicative of strong $x$-$\Phi$ correlations in
the source.

\section{Azimuthal analysis of the fragmentation region}
\label{sec5}

For finite impact parameter, the collision system has a
finite total angular momentum $\vec{L}$,  
$L_i = \epsilon_{ijk} \langle\!\langle x_j\, p_k\, \rangle\!\rangle$.
In heavy ion collisions, $\vec{L}$ and $L_{\rm tot} = \vert\vec{L}\vert$,  
is not directly observable but has to
be infered indirectly on the basis of impact parameter dependent
measurable quantities like total particle multiplicities or transverse 
energy. In general, angular momentum conservation is a constraint 
which can affect the shape of two-particle correlations~\cite{KBS89}.

For the incoming nuclei, the angular
momentum is entirely determined by $L_{\rm tot} = \langle\!\langle x\, p_z\, 
\rangle\!\rangle$, but in the final state, the component 
$\langle\!\langle z\, p_x\, \rangle\!\rangle$ in general 
carries part of $L_{\rm tot}$. To illustrate the consequences, we consider 
a longitudinally expanding collision scenario, for which
the space-time rapidity of the emission points 
$\eta = \textstyle{1\over 2} \log\textstyle{ {t-z}\over {t+z}}$
is linearly related to the momentum rapidity $Y$ of the emitted particles,
$\eta = \eta_l\cdot Y$. 
Using $z = \tau \sinh\eta = \tau \sinh(\eta_l Y)$, one sees that a 
non-vanishing contribution $\langle\!\langle z\, p_x\, \rangle\!\rangle$
to the total angular momentum leads to a non-vanishing total vector sum 
of $p_x$ in non-central rapidity bins, and hence to a non-vanishing 
Fourier coefficient $v_1$ in (\ref{1.1}). This effect is more enhanced 
for larger rapidities $Y$,
it vanishes at mid-rapidity $Y = 0$ and it has opposite sign
for backward rapidities. 
The size of the fraction 
$\langle\!\langle z\, p_x\, \rangle\!\rangle$ of $L_{\rm tot}$, 
depends on details of the dynamics. Note however,
that $\langle\!\langle z\, p_x\, \rangle\!\rangle \not= 0$ does not
imply necessarily the equilibration of microscopic degrees of freedom or 
source gradients.

According to this simplified scenario, the $180^o$-symmetry
of the transverse collision region is lost at non-central rapidities,
and the odd harmonic coefficients of the one- and two-particle
spectra do not have to vanish any more. In this section, we 
investigate the implications for the first harmonics of the
HBT radius parameters.

\subsection{Properties of first harmonics}
\label{sec51}

To calculate odd harmonic coefficients of the transverse
HBT radius parameters (\ref{2.3}), we start again from
the harmonic expansion (\ref{3.4}) of $T^\perp$.
It is a remarkable property that in the
radii ${R_{ij,m}^c}^2$, ${R_{ij,m}^s}^2$, even and odd harmonic 
coefficients $A_n$, ${A'}_n$, etc. do not mix. For $m$ even, only
terms with $n$ even appear, and for $m$ odd, only terms with
$n$ odd. Especially, we find for the first harmonic coefficients
  \begin{mathletters}
    \label{5.1}
  \begin{eqnarray}
     {R_{s,1}^c}^2 &=& \textstyle{1\over 2}A_1
                       - \textstyle{1\over 4}B_1 
                       - \textstyle{1\over 4}C_1 
                       - \textstyle{1\over 4}B_3 
                       - \textstyle{1\over 4}C_3 \, ,
     \label{5.1a} \\
     {R_{o,1}^c}^2 &=&  \textstyle{1\over 2}A_1
                       + \textstyle{1\over 4}B_1 
                       + \textstyle{1\over 4}C_1 
                       + \textstyle{1\over 4}B_3 
                       + \textstyle{1\over 4}C_3 \, ,
     \label{5.1b} \\
     {R_{os,1}^s}^2 &=&  - \textstyle{1\over 4}B_1 
                       - \textstyle{1\over 4}C_1 
                       + \textstyle{1\over 4}B_3 
                       + \textstyle{1\over 4}C_3 \, ,
     \label{5.1c} \\
     {R_{s,1}^s}^2 &=& {R_{o,1}^s}^2 = {R_{os,1}^c}^2 = 0\, .
  \end{eqnarray}
  \end{mathletters}
It follows immediately that 
all first harmonics vanish in the absence of source gradients,
  \begin{equation}
    {R_{ij,1}^c}^2 = {R_{ij,1}^s}^2 = 0\, ,
    \qquad \hbox{[no source gradients].}
    \label{5.3}
  \end{equation}
The physical
reason is that for a source without gradients, the effective emission
region has the same {\it side} and {\it out} extension irrespective 
of whether it is viewed under an angle $\Phi$ or an angle 
$\Phi + \pi$.

For the case that all third harmonic coefficients
in (\ref{5.1}) are negligible, the three non-vanishing HBT radii in 
(\ref{5.1}) satisfy
  \begin{equation}
    {R_{s,1}^c}^2 \approx {R_{o,1}^c}^2 + 2{R_{os,1}^s}^2\, .
    \label{5.5}
  \end{equation}
This equation is reminiscent of (\ref{3.9}). There, however,
the contribution $B_0$ is typically an order of
magnitude larger than the second harmonic contribution $A_2$,
and this allows for the further simplification (\ref{3.3b}).
Here, in contrast, all leading terms are first harmonics.
The first harmonic coefficient of $\langle \tilde{x}^2\rangle$ 
should be much larger than that of 
$\langle \tilde{y}^2\rangle$ since asymmetries with respect
to the beam axis will occur in the direction of the impact
parameter only. In this limiting case which is relevant
for the models studied below, we have $A_1 = B_1 \gg C_1$,
and 
  \begin{equation}
        {R_{o,1}^c}^2\, :\, {R_{s,1}^c}^2\, :\,  {R_{os,1}^s}^2\, 
        \approx \, 3\, :\, 1\, :\, -1\, .
    \label{5.6}
  \end{equation}

\subsection{Model extensions for non-central rapidities}
\label{sec52}

The model emission function (\ref{4.1}) investigated in section~\ref{sec4}
describes a collision with vanishing angular momentum. 
Here, we introduce a simple extension which allows for
finite angular momentum $L_{\rm tot}$ but coincides with the model
(\ref{4.1}) at mid-rapidity. Clearly, such extensions are numerous. One can
e.g. shift the transverse flow pattern in the $x$-direction, 
  \begin{equation}
    u_x^\chi = \eta_f \sqrt{1+\epsilon_f} { {x+\chi Y}\over R}\, .
    \label{5.7}
  \end{equation}
Since $\chi$ multiplies the rapidity $Y$, the flow pattern shows a 
forward-backward anticorrelation in $Y$ and will result in non-vanishing
odd coefficients $v_n$ and a finite
total angular momentum (\ref{1.4}). Alternatively, one can introduce a
rapidity-dependent deformation of the transverse geometry, e.g. by
introducing a dependence of the Gaussian widths in (\ref{4.2}) on 
the azimuthal direction $\varphi$, $x = r\, \cos\varphi$
  \begin{eqnarray}
    \rho_x &=& \left(R+\chi Y\cos\varphi\right)\, \sqrt{1-\epsilon_s}\, ,
    \nonumber \\
    \rho_y &=& \left(R+\chi Y\cos\varphi\right)\, \sqrt{1+\epsilon_s}\, .
    \label{5.8}
  \end{eqnarray}
For finite transverse flow, this again results in non-vanishing
odd coefficients $v_n$ and a finite $L_{\rm tot}$.
Both these modifications, (\ref{5.7}) and (\ref{5.8}) reduce at
mid-rapidity $Y = 0$ to the model studied in section~\ref{sec4}. 
We have investigated the $K_\perp$-dependence
of their first and second harmonics and the (approximate) relations which
they satisfy. The linear coupling of (\ref{5.7}) on $\Phi$-dependent
terms in (\ref{4.6}) makes the $\chi$-dependence of the model 
(\ref{5.7}) very weak. (The term $K_\mu {u_x^\chi}^\mu$ does not 
introduce an additional position-momentum correlation,
and the only $\chi$-dependence of the correlator stems from $u_l$).
Here, we present results for the model (\ref{5.8})
only. To isolate the effect of the $\chi$-displacement, we have set 
the other anisotropy parameters to zero, $\epsilon_s = \epsilon_f = 0$.
All calculations are done at forward rapidity, $Y = 1$.

\begin{figure}[h]\epsfxsize=8cm 
\centerline{\epsfbox{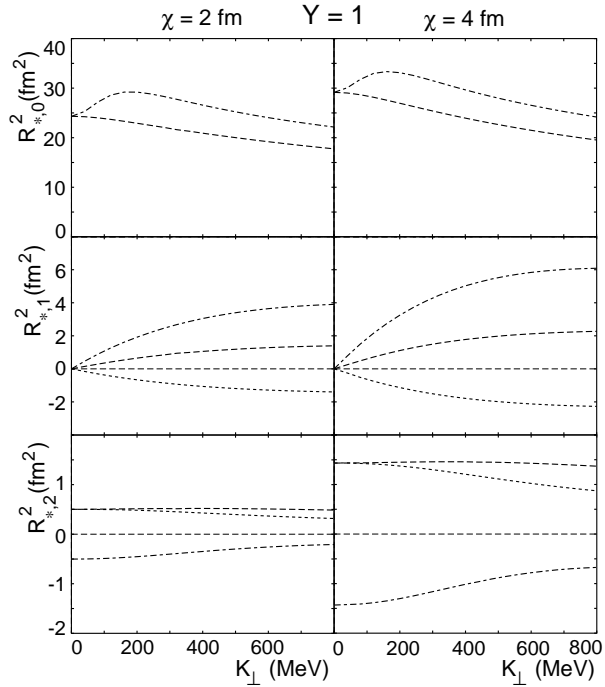}}
\caption{Zeroth, first and second harmonic coefficients of the
out (dash-dotted lines), side (dashed lines) and out-side (dotted
lines) HBT radius parameters for the model (\protect{\ref{4.1}}) 
with the modification (\protect{\ref{5.8}}) at forward
rapidity $Y=1$. To high accuracy,
the first harmonic coefficients satisfy the relation 
(\protect{\ref{5.6}}) and the second harmonics the relation 
(\protect{\ref{3.9}}).
}\label{fig6}
\end{figure}

For vanishing transverse flow, $\eta_f = 0$, our model contains no
source gradients. The zeroth harmonic coefficients show the expected
behaviour: $R_{s,0}$ is a $K_\perp$-independent constant from
which the out radius $R_{o,0}$ differs by the factor 
$\beta_\perp^2 \langle\tilde{t}^2\rangle$ only. Also, in
the absence of source gradients, all first harmonics vanish, and
(\ref{5.3}) holds. The
second harmonics do not vanish: they are $K_\perp$-independent
parameters determining the elliptic approximation of the transverse 
source geometry. Also, they satisfy the geometrical
relation (\ref{3.3b}) as expected for sources without 
position-momentum correlations. 
Qualitatively, their behaviour is completely consistent with the case
discussed for $\eta_f = 0$ in Fig.~\ref{fig4}. Quantitatively, we find 
for the non-vanishing components $\vert {R_{ij,2}^*}^2\vert$ $\approx 0.5$
if $\chi = 2$ fm and $\vert {R_{ij,2}^*}^2\vert$ $\approx 1.5$ 
if $\chi = 2$. This is slightly
larger than the values obtained for the case $\eta_f = 0.3$ at 
$K_\perp = 0$ (see Fig.~\ref{fig6}) and the minimal $\eta_f$-dependence
can be traced back to the $u_l$-dependent term in the Boltzmann factor
(\ref{4.6}).

For finite transverse flow $\eta_f = 0.3$, 
numerical results are presented in Fig.~\ref{fig6}. Due to source
gradients, the azimuthal eccentricity introduced via (\ref{5.8})
now shows up in the first harmonic coefficients. They
vanish at vanishing transverse pair momentum,
  \begin{equation}
    \lim_{K_\perp\to 0} {R_{ij,1}^c}^2(K_\perp,Y)
    = \lim_{K_\perp\to 0} {R_{ij,1}^s}^2(K_\perp,Y) = 0\, ,
    \label{5.9}
  \end{equation}
since the correlator is at $K_\perp = 0$ sensitive to the geometry
of the source only. More importantly, over the complete 
$K_\perp$-range, the first harmonics in the out, side and
out-side directions satisfy up to a few percent the relation
$3\, : \, 1 \, : -1$ as we have argued in deriving (\ref{5.6}).
The second harmonic coefficients satisfy for $K_\perp = 0$
the relation $-1\, :\, 1\, : \, 1$ as suggested by (\ref{3.3b}).
For finite $K_\perp$, dynamically introduced deviations are 
found, but the modified relation (\ref{3.9}),
${R_{o,2}^c}^2 +$ $2{R_{os,2}^s}^2$ $= {R_{s,2}^c}^2$
describes all results up to a few percent. 
%
%
\section{Finite event statistics and finite multiplicity fluctuations}
\label{sec6}

For the particle multiplicities obtained at the AGS and SPS, a 
determination of HBT radius parameters on an event-by-event basis
is not possible. Finite event statistics limits the 
possibilities of a multidimensional HBT analysis for the now typical
samples of the order of $10^5$ - $10^6$ events. To extract despite these
statistical constraints at least the major anisotropic HBT characteristics,
it is clearly helpful to start from an azimuthal parametrization of 
$C({\bf K},{\bf q})$ in terms of a minimal set of fit parameters.
Such a parametrization is discussed in section~\ref{sec6a}. 

In contrast to constraints from finite event statistics, the statistical
uncertainties stemming from finite multiplicity fluctuations cannot 
be overcome by investigating larger event samples. They constitute a
fundamental limitation to any investigation of anisotropy measures,
and they are particularly important in reconstructing the 
reaction plane. In section~\ref{sec62} we investigate to what extent
this affects the determination of anisotropy measures from two-particle
correlators.

\subsection{A minimal azimuthal parametrization}
\label{sec6a}

On the basis of our discussion in the sections~\ref{sec2} - \ref{sec5},
we propose as minimal parametrization of a two-particle
correlator for peripheral collisions the Gaussian ansatz
  \begin{mathletters}
    \label{6.1}
  \begin{eqnarray}
    C_{\psi_R}({\bf K},{\bf q}) &\approx& 1 + 
        \lambda\, \,  C_{sym}({\bf K},{\bf q})\,
                      C_1({\bf K},{\bf q},\psi_R)\,
        \nonumber \\
        && \qquad \qquad \times
                C_2({\bf K},{\bf q},\psi_R)\, 
        \label{6.1a} \\
    C_{sym}({\bf K},{\bf q}) &=& \exp\lbrack
        -{R_{o,0}}^2\, q_o^2 - {R_{s,0}}^2\, q_s^2
        \nonumber \\
             && \qquad
             - {R_{l,0}}^2\, q_l^2 - 2\, {R_{ol,0}}^2\, q_o q_l \rbrack\,
        \label{6.1b} \\
    C_1({\bf K},{\bf q},\psi_R)    
        &=& \exp\big\lbrack -\alpha_1\,
                (3\, q_o^2 + q_s^2)\, \cos(\Phi-\psi_R)\,
        \nonumber \\
        && \qquad
                + 2\alpha_1\, q_o q_s \sin(\Phi -\psi_R) \big\rbrack
        \label{6.1c} \\
    C_2({\bf K},{\bf q},\psi_R)
        &=& \exp\big\lbrack -\alpha_2
                (q_o^2-q_s^2)\cos 2(\Phi-\psi_R)
        \nonumber \\
        && \qquad
                +2\,\alpha_2 q_oq_s \sin 2(\Phi-\psi_R) \big\rbrack\, .
    \label{6.1d}
  \end{eqnarray}
  \end{mathletters}
The zeroth harmonic coefficients in this parametrization provide 
the complete Cartesian parametrization for the azimuthally symmetric 
case. Especially, the cross term ${R_{ol,0}}^2$ vanishes 
in the CMS at mid rapidity, and due to the symmetry arguments made 
in section~\ref{sec4}, the parameter $\alpha_1$ vanishes under
these conditions as well. The parametrization for 
the first effective harmonic coefficient
$\alpha_1$ and the second one $\alpha_2$ is motivated by the
relations (\ref{5.6}) and (\ref{3.3b}), i.e., it is based on setting
  \begin{mathletters}
    \label{6.2}
  \begin{eqnarray}
    \alpha_1(K_\perp,Y) \approx {R_{s,1}^c}^2 
                        \approx \textstyle{1\over 3} {R_{o,1}^c}^2 
                        \approx -{R_{os,1}^s}^2 \, ,
    \label{6.2a} \\
    \alpha_2(K_\perp,Y) \approx {R_{o,2}^c}^2 
                        \approx - {R_{s,2}^c}^2 
                        \approx - {R_{os,2}^s}^2\, . 
    \label{6.2b}
  \end{eqnarray}
  \end{mathletters}
In the model studies in section~\ref{sec4} and \ref{sec5}, we have
assumed that $\Phi = 0$ corresponds to the  pair momentum
${\bf K}$ lying in the reaction plane. Experimentally, this 
reaction plane is unknown a priori. The parameter $\psi_R$
which determines its orientation has to be included 
in a comparison with experiment. A discussion of the statistical
uncertainty in its determination and the implication for the
extraction of the anisotropy parameters $\alpha_1$, $\alpha_2$
is given in section~\ref{sec62}.

In the model-independent analysis of the $\Phi$-dependent HBT radius
parameters in sections~\ref{sec3} and \ref{sec4}, we have argued 
that corrections to these relations can be expected to be comparatively 
small on general grounds. Also, we have quantified deviations from 
(\ref{6.2}) in the model studies of section~\ref{sec4} and ~\ref{sec5}. 
According to these studies, the most reasonable non-minimal extension of 
the parametrization (\ref{6.1}) is to implement the constraint
${R_{o,2}^c}^2$ $+ 2{R_{os,2}^s}^2$ $\approx$ ${R_{s,2}^c}^2$ 
instead of (\ref{3.3b}), i.e., to replace (\ref{6.2b}) by two
parameters for the second harmonics.

Clearly, it would be preferable to 
fit to the most general parametrization (\ref{1.3}), and to quantify 
corrections to these relations. However, as long as finite event 
statistics forces one to restrict the space of fit parameters as far 
as possible, the relations (\ref{6.2}) provide in the light of our
analysis the most reasonable set of constraints to adopt. 

\subsection{Event samples with oriented reaction plane}
\label{sec62}

The main problem in extracting the fit parameters $\alpha_1$, $\alpha_2$
from experimental data is, that due to finite event multiplicity,
a multidimensional analysis of two-particle correlations is not
possible on an event-by-event basis. Fits have to be done on
event samples and the use of (\ref{6.1}) presupposes that the
different events are sampled with a fixed orientation of the
reaction plane. The reaction plane can be reconstructed from an
azimuthal analysis of the single particle distribution (\ref{1.1}),
but its event-by-event determination is subject to significant  
statistical uncertainties. Since the parametrization (\ref{6.1})
depends on the angle $\psi_R$, these
uncertainties affect the determination of the HBT radius parameters
as well. Here, we investigate quantitatively to what extent
these uncertainties affect the determination of the anisotropy
parameters $\alpha_1$, $\alpha_2$ from the experimental data.

Our starting point is the assumption that the probability
distributions $W(v_1,\psi_R)$ of the first harmonic coefficients
of the one-particle spectrum around $(\bar{v}_1,\bar{\psi}_R)$
is given by~\cite{O92,O93,VZ96}
  \begin{equation}
    W(v_1,\psi_R) = {1\over 2\pi\sigma^2}
    \exp\left(-{ {\bar{v}_1^2 + v^2_1 - 2\bar{v}_1v_1\, 
          \cos \psi_R} \over 2\sigma^2} \right)\, .
    \label{6.3}
  \end{equation}
We consider the case that the reaction plane is reconstructed 
from the first harmonic coefficients only, and we use the
shorthand $\psi_R = \psi_R^{(1)}$. For ideas about  
improving this reconstruction by taking higher order harmonics
$(v_i,\psi_R^{(i)})$ into account as well, we refer to~\cite{VZ96}. 
Here and in what follows, we orient the most likely, `true' direction
along the $x$-axis, $\bar{\psi}_R = 0$.
The Gaussian distribution (\ref{6.3}) can be used if 
event multiplicities $N$ are sufficiently large to apply the central
limit theorem. The variance $\sigma^2$ then scales like 
$\textstyle{1\over N}$, and the reaction plane is well
defined for extremely large event multiplicities,
  \begin{equation}
    \lim_{N\to \infty}\, W(v_1,\psi_R) 
    = \delta\left( \psi_R \right)\, .
    \label{6.4}
  \end{equation}
For finite multiplicities $N$ in the hundreds, however, the 
uncertainty in the eventwise determination of $\bar{\psi}_R$ 
cannot be neglected. The
event sample with oriented reaction plane should not be compared
directly to (\ref{6.1}), but to an effective correlator which
takes the probability distribution $W$ of experimentally
determined reaction plane orientations properly into account, 
  \begin{equation}
    C_{\bar{\psi}_R}^{\rm eff}({\bf K},{\bf q})
    = \int v_1\, dv_1\, d\psi_R\, W(v_1,\psi_R)\,
    C_{{\psi}_R}({\bf K},{\bf q})\, .
    \label{6.5}
  \end{equation}
It follows from the form of (\ref{6.3}) that
this effective correlator does not depend on $\bar{v}_1$ and
$\sigma$ separately, but is a function of 
$\bar{\xi} = \bar{v}_1/\sigma$ only. The parameter $\bar{\xi}$
is a direct measure of the accuracy for the reaction plane
orientation~\cite{O92,O93,VZ96}. From the investigation of Voloshin and
Zhang (see Fig.~4 in~\cite{VZ96}), we conclude that a value of 
$\bar{\xi} \approx 2$ corresponds to an uncertainty of approximately $30^o$.
This is the current experimental standard which will improve with 
the larger event multiplicities at RHIC and LHC. 

\begin{figure}[h]\epsfxsize=7cm 
\centerline{\epsfbox{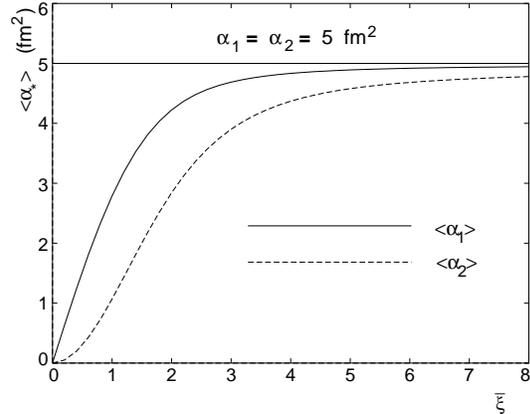}}
\caption{The HBT anisotropy parameters $\langle\alpha_i\rangle$ as a function
of the parameter $\bar{\xi}$ which characterizes the event-by-event
reconstruction uncertainty in the orientation of the reaction plane.
The parameters $\langle\alpha_1\rangle$, $\langle\alpha_2\rangle$ 
are determined by fitting (\protect{\ref{6.1}}) to an event sample 
(\protect{\ref{6.5}}) of correlators whose reaction planes
are oriented along the $x$-axis according to the probability
distribution (\protect{\ref{6.3}}). The value $\bar{\xi} = 2$
corresponds to a reconstruction uncertainty of approximately $30^o$.
}\label{fig7}
\end{figure}

We investigate now in a simplified example to what extent finite 
anisotropies of the single events leave traces in an event sample 
constructed as outlined above. To this aim, we calculate 
for given anisotropies $\alpha_1$, $\alpha_2$ the effective 
correlator $C_{\bar{\psi}_R}^{\rm eff}$ which is then 
fitted to the Gaussian ansatz (\ref{6.1}).
The anisotropy parameters $\langle \alpha_1\rangle$,
$\langle \alpha_2\rangle$ determined in this fit are then 
compared to the parameters $\alpha_1$, $\alpha_2$ of the 
single events one has started with. In Fig.~\ref{fig7}
the fitted averages $\langle \alpha_i\rangle$ are shown as 
function of the statistical uncertainty $\bar{\xi}$ in the 
reconstruction of the reaction plane from first harmonic 
coefficients. For a realistic uncertainty of $30^o$ 
($\bar{\xi} \approx 2$), approximately 85 \% of the `true' 
first anisotropy parameters $\alpha_1$ and approximately
55 \% of the second anisotropy parameter $\alpha_2$  
is obtained. These ratios are independent of the size 
of $\alpha_1$ and $\alpha_2$. 

The correlator of an unoriented event sample is obtained from
(\ref{6.1}) by averaging $C_{\bar{\psi}_R}$ over all orientations
of the reaction plane. It is known that in fits to such 
`unoriented' correlators, HBT radius parameters 
receive artificial contributions due to the averaging 
procedure~\cite{VC96b}. Since we average in $C_{\bar{\psi}_R}^{eff}$ 
over different reaction planes, such artificial contributions 
exist for $C_{\bar{\psi}_R}^{\rm eff}$ too. Especially, the zeroth 
harmonic coefficients of the HBT radius parameters should show a
$\bar{\xi}$-dependence. In the numerical analysis of
the curvature of the correlator, we found this to be negligible.
To understand the reason, we consider the case $\alpha_2 = 0$
when the integral (\ref{6.5}) can be evaluated analytically.
The $\psi_R$-dependent part reads 
  \begin{eqnarray}
     C^{\rm eff}_1({\bf K},{\bf q}) &=& 
     \int v_1\, dv_1\, d\psi_R\, W(v_1,\psi_R)\, 
         C_1({\bf K},{\bf q},\psi_R)\, 
     \nonumber \\
     &=& \int \xi\, d\xi\, 
         \exp\left( -\textstyle{1\over 2}
                \left( \xi^2 + \bar{\xi}^2\right) \right)\, 
              I_0(Z)\, .
     \label{6.6}
  \end{eqnarray}
The argument $Z$ of the Bessel function
depends on the relative pair momentum components $q_o$, $q_s$
and the anisotropy parameters $\alpha_1$
  \begin{mathletters}
    \label{6.7}
  \begin{eqnarray}
    Z &=& \left(\bar{\xi}\xi\right)^2\,  
            - 2\bar{\xi}\xi 
             \left(F_c\cos(\Phi) + F_s\sin(\Phi)\right)
            \nonumber \\
            && \qquad
               + 4\, F_c\, F_s\, \sin(\Phi)\, \cos(\Phi)\, 
               + {F_c}^2\, + {F_s}^2\, ,
        \label{6.7a} \\
    F_c &=& \alpha_1 \left(3\, q_o^2 + q_s^2\right)\, ,\qquad
    F_s = -2\, \alpha_1\, q_o\, q_s\, .
        \label{6.7b}
  \end{eqnarray}
  \end{mathletters}
The limit $\bar{\xi} \to 0$ in (\ref{6.5}) corresponds to an
unoriented correlator. For this case, the argument $Z$ is quadratic in 
the components of ${\bf q}$ and expanding $I_0$ for small
arguments, we find, (\ref{a5}), 
  \begin{equation}
    \left. { \partial^2 I_0(Z)\over 
      \partial q_i\, \partial q_j}
        \right\vert_{ {\bf q} = 0} = 0\, \qquad
        \hbox{for $\bar{\xi} = 0$.}
  \end{equation}
This illustrates for a simple example that the unweighted averaging
over different event orientations discussed in \cite{VC96b} does
affect the shape of the correlator in ${\bf q}$, but {\it not}
its curvature. It hence has to be determined by quantifying 
the deviations of $C^{\rm eff}_{\bar{\psi}_R}$ from a Gaussian
shape~\cite{WH96}. Here, we do not pursue this point further.
%
%
The calculation leading to Fig.~\ref{fig7} is based on simplified
assumptions. Especially, we have not considered eventwise 
fluctuations in the size of the $\alpha_i$. Since both anisotropy fit
parameter can take positive and negative values, 
such fluctuations do not fake anisotropy signals. We hence
conclude from the above that with a typical $30^o$ resolution
of the reaction plane orientation, a determination of the anisotropy
parameters $\alpha_1$, $\alpha_2$ from experimental data is
feasible.
%
\section{Conclusion}
\label{sec7}

In the present work, we have studied the possibilities of an harmonic
analysis of two-particle HBT radius parameters. We have clarified to
what extent the two fundamental problems of such an analysis can
be overcome:

Firstly, the harmonic Fourier expansion of HBT radius parameters 
introduces a plethora of additional fit parameters which make a
comparison with experimental data difficult. We have argued
that under reasonable assumptions on the geometry and dynamics of the
collision region, only very few of them are non-negligible. For
the first harmonic coefficients, these are ${R_{o,1}^c}^2$,
${R_{s,1}^c}^2$, ${R_{os,1}^s}^2$, and they can be described by one
single parameter since their leading contributions scale like
$3\, :\, 1\, : \, -1$. Amongst the second harmonic coefficients, 
only ${R_{o,2}^c}^2$, ${R_{s,2}^c}^2$, ${R_{os,2}^s}^2$ are 
non-negligible and their leading contributions scale like
$-1\, :\, 1\, : \, 1$. Higher harmonic coefficients were found in 
all studies to be an order of magnitude smaller.
Based on these observations, our main
result is the parametrization (\ref{6.1}) which describes the
leading anisotropy of the two-particle correlator by two
additional fit parameters only. If deviations from this 
parametrization turn out to be important, than this will 
provide an important constraint on further model studies. 
A first non-minimal parametrization suggested by our studies
quantifies deviations of the second harmonic coefficients
from $-1\, :\, 1\, : \, 1$ by one additional fit parameter.

The second fundamental problem in the determination of anisotropy
signals from $C({\bf K},{\bf q})$ is the statistical uncertainty
in the eventwise reconstruction of the reaction plane. We have
shown that for event samples with a typical $30^o$ uncertainty
in the eventwise orientation, the major part of the anisotropy
measures $\alpha_1$, $\alpha_2$ survives. An analysis of 
experimental data on the basis of the parametrization (\ref{6.1})
seems hence feasible.

We finally discuss which physical information can be
extracted from the anisotropy parameters $\alpha_1$, $\alpha_2$.
From the discussion in sections~\ref{sec3} - \ref{sec5}, we 
conclude that a non-vanishing first harmonic coefficient $\alpha_1$
automatically implies the existence of dynamical source gradients, see
Eq. (\ref{5.3}). In contrast, the leading contribution to
$\alpha_2$ is determined by the geometry of the collision region
and specifies essentially the elliptic shape of its transverse
extension. More explicitly, in the context of the hydrodynamical 
model of section~\ref{sec3}, there is a tentative strategy to
determine the geometrical and dynamical anisotropies $\epsilon_s$,
$\epsilon_f$ respectively. According to (\ref{4.8}),
experimental data on the harmonic coefficient $v_2$ restricts the
allowed parameter space in $(\epsilon_s,\epsilon_f)$ to a 
one-dimensional one. According to (\ref{4.10a}), the second harmonic 
coefficients of the transverse HBT radii show a different dependence
on $\epsilon_s$ and $\epsilon_f$ and can hence be used to constrain
the remaining freedom. This illustrates that in non-central
collisions, as in azimuthally symmetric ones, only a combination
of 1- and 2-particle spectra will allow to disentangle geometrical
and dynamical information.

\acknowledgements
The author thanks U. Heinz for helpful discussions and a critical reading
of the manuscript. Also, discussions with P. Braun-Munziger and 
S. Voloshin at an early stage of this project are gratefully
acknowledged. This work was supported by BMBF and DFG.

\appendix
\section{Calculation of azimuthal particle distributions and HBT
radius parameters}
\label{appa}

In this appendix, we give details of how to calculate the harmonic
coefficients of the single particle distributions (\ref{1.1}) and
the two-particle correlations (\ref{1.3}) for the model in
section~\ref{sec4},
  \begin{eqnarray}
    v_n &=& \int_0^{2\pi} d\phi\, \cos n\phi\, \int d^4x\, S(x,K)\,
    \nonumber \\
    &=& \tau_0 m_\perp\, \int {\cal D}\eta\, G_n(\eta)\, ,
    \label{a1} \\
    \int {\cal D}\eta\, &=& \int \tau\, d\tau\, d\eta\, \cosh(\eta-Y)\,
        {\rm e}^{ -\textstyle{ {(\tau-\tau_0)^2}\over {2(\Delta\tau)^2}}
                  -\textstyle{ {\eta^2}\over {2(\Delta\eta)^2}}}\, .
    \label{a2}
  \end{eqnarray}
For $n$ odd, all coefficients $v_n$ vanish due to the 
$180^o$-symmetry in the transverse plane. 
In the approximation $u_l \approx 1 + \textstyle{1\over 2}u_x^2
+ \textstyle{1\over 2}u_y^2$, the Boltzmann term $K^\mu u_\mu$ 
is quadratic in $x$ and $y$, and 
  \begin{eqnarray}
    G_n(\eta) &=& \int_0^{2\pi} d\phi\, \cos n\phi\, \int dx\, dy\,
    {\rm e}^{ -\textstyle{ {K^\mu u_\mu}\over T}
              -\textstyle{ {x^2}\over {2\rho_x^2}}
              -\textstyle{ {y^2}\over {2\rho_y^2}} }
            \nonumber \\
        &=& 2\pi^2\lambda_x\lambda_y\sqrt{g_xg_y}
            {\rm e}^{-A + \textstyle{ {K_\perp^2}\over 2T^2}
                       \textstyle{ {g_x+g_y}\over 2} }
              I_{n\over 2}(Z)\, ,
    \label{a3}
  \end{eqnarray}
where we have used
  \begin{eqnarray}
    g_x &=& 1/ \left(A + {\lambda_x^2\over \rho_x^2}\right) \, ,\qquad
    A = \textstyle{m_\perp\over T}\cosh(\eta-Y)\, ,
    \nonumber \\
    Z &=& \left( \textstyle{ {K_\perp^2}\over 2T^2}
                          \textstyle{ {g_x-g_y}\over 2} \right)
       = {{K_\perp^2}\over 2T^2}
         {{\left( {\lambda_y^2\over \rho_y^2}
                     - {\lambda_x^2\over \rho_x^2} \right)}\over
          { (A + {\lambda_x^2\over \rho_x^2})\, 
            (A + {\lambda_y^2\over \rho_y^2})} }\, .
    \label{a4}
  \end{eqnarray}
The modified Bessel function $I_{n\over 2}$ in (\ref{a3}) is obtained
by doing the $\phi$-integral. To extract the leading dependence in
$Z$, we expand for small arguments,
  \begin{eqnarray}
    I_{n\over 2}(Z) &=& \textstyle{1\over 2\pi} 
         \int_0^{2\pi} d\phi\, \cos n\phi\, {\rm e}^{Z\cos 2\phi}\,
    \nonumber \\
    &=& \sum_{k=0}^\infty {1\over k! \Gamma(\textstyle{n\over 2}+k+1)}
        \left(Z\over 2\right)^{\textstyle{n\over 2}+2k}\, .
       \label{a5}
  \end{eqnarray}
For the coefficient $n = 2$, the leading $Z$-dependence in (\ref{a4}) is
linear. This implies (\ref{4.8}). Also, for small 
$K_\perp$, $A$ is approximately constant and $Z \propto K_\perp^2$.
Since $A$ depends on $\eta$, the $\eta$-integration in (\ref{a1})
has to be done numerically. To obtain simple
expressions for the main qualitative features, one may use
  \begin{equation}
    A \approx \textstyle{m_\perp\over T}\, .
    \label{a6}
  \end{equation}
This amounts to $\cosh(\eta - Y) \approx 1$ and
allows for the approximate analytical calculation of HBT radius 
parameters. The latter are calculated via space-time variances which
we express here in terms of the averages
  \begin{equation}
    \langle{f(x,y)}\rangle_* = \int dx\, dy\, f(x,y)\, 
    {\rm e}^{ -\textstyle{ {K^\mu u_\mu}\over T}
              -\textstyle{ {x^2}\over {2\rho_x^2}}
              -\textstyle{ {y^2}\over {2\rho_y^2}} }\, ,
    \label{b1}
  \end{equation}
  \begin{equation}
    \langle x_\mu x_\nu \rangle
     = { \int {{\cal D}\eta \langle{x_\mu x_\nu}\rangle_*}
                  \over {\int {\cal D}\eta \langle{1}\rangle_*}}\, .
    \label{b2}
  \end{equation}
Again, in the approximation $u_l \approx 1 + \textstyle{1\over 2}u_x^2
+ \textstyle{1\over 2}u_y^2$, the Boltzmann term $K^\mu u_\mu$ is 
quadratic in $x$ and $y$, which allows for an analytical calculation
of the $x$- and $y$-integration. We find $\langle x\, y\rangle_* = 0$, and
  \begin{eqnarray}
    {\langle x^2\rangle_*}
      &=& {\lambda_x^2\, \langle 1\rangle_*^{-1}  \over 
      {A + \textstyle{\rho_x^2\over \lambda_x^2}} }
      = { R^2\, (1-\epsilon_s)\, \langle 1\rangle_*^{-1}\over
          {1 + A\eta_f^2 (1-\epsilon_s)\, (1+\epsilon_f)}}\, ,
      \nonumber \\
    {\langle y^2\rangle_*} 
      &=& {\lambda_y^2 \, \langle 1\rangle_*^{-1} \over 
      {A + \textstyle{\rho_y^2\over \lambda_y^2}} }
      = { R^2\, (1+\epsilon_s)\, \langle 1\rangle_*^{-1}\over
          {1 + A\eta_f^2 (1+\epsilon_s)\, (1-\epsilon_f)}}\, .
     \label{b3}
  \end{eqnarray}
These averages are the building blocks for the space-time
variance (\ref{b2}) which determine the HBT radius parameters. Since $A$ 
depends on $\eta$, the remaining $\eta$-integration has to be done 
numerically. With the help of the approximation (\ref{a6}), however,
the $\eta$-integration in (\ref{b2}) drops out and the analytical 
expressions (\ref{4.10}) for $\bar{g}$ and $\bar{g}\alpha_T$ can
be obtained. Their validity is subject to the approximations made
above but they give a qualitatively correct, simple and intuitive 
description of the numerical results. Most remarkably, these 
approximate expressions do not depend on the azimuthal direction 
of ${\bf K}$, cf. the discussion in section \ref{sec42}.

%


\begin{references}
\bibitem{comment}
For an azimuthally symmetric model with 
quadratic flow profile, the $K_\perp$-dependence of $R_o$ was
calculated already in ~\cite{WSH96}. In contrast to Fig.~\ref{fig5},
a positive slope was found above $K_\perp \approx 400$ MeV. This
can be traced back to using a transverse flow profile
$\eta_f \textstyle{x^2\over R^2}$ which unlike (\ref{4.12}) has 
an unrealistic shape for $x < 0$. Concerning this point, the
discussion in \cite{WSH96} is misleading.
\end{references}
\end{document}